\documentclass[ reprint, aps, prl, two column, showkeys, groupedaddress]{revtex4-1}
\usepackage{graphicx,epsfig}
\usepackage{amssymb}
\usepackage{amsmath}
\usepackage{bm}
\usepackage{textcomp}
\usepackage{soul,xcolor}

\usepackage{color}

\begin{document}
\setcounter{page}{1}
\setstcolor{red}

\title[]{Multivalley two-dimensional electron system in an AlAs quantum well with mobility exceeding $2\times10^6$ cm$^{2}$V$^{-1}$s$^{-1}$}
\author{Yoon Jang \surname{Chung}}
\author{K. A. \surname{Villegas Rosales}}
\author{H. \surname{Deng}}
\author{K. W. \surname{Baldwin}}
\author{K. W. \surname{West}}
\author{M. \surname{Shayegan}}
\author{L. N. \surname{Pfeiffer}}
\affiliation{Department of Electrical Engineering, Princeton University, Princeton, NJ 08544, USA  }
\date{\today}

\begin{abstract}

Degenerate conduction-band minima, or `valleys', in materials such as Si, AlAs, graphene, and MoS$_2$ allow them to host two-dimensional electron systems (2DESs) that can access a valley degree of freedom. These multivalley 2DESs present exciting opportunities for both pragmatic and fundamental research alike because not only are they a platform for valleytronic devices, but they also provide a tool to tune and investigate the properties of complex many-body ground states. Here, we report ultra-high quality, modulation doped AlAs quantum wells containing 2DESs that occupy two anisotropic valleys and have electron mobilities  peaking at $2.4\times10^6$ cm$^{2}$V$^{-1}$s$^{-1}$ at a density of $2.2\times10^{11}$ cm$^{-2}$. This is more than an order of magnitude improvement in mobility over previous results. The unprecedented quality of our samples is demonstrated by magneto-transport data that show high-order fractional quantum Hall minima up to the Landau level filling $\nu=8/17$, and even the elusive $\nu=1/5$ quantum Hall state.

%, surrounded by insulating phases which are generally interpreted as pinned Wigner crystal states.

\end{abstract}
\maketitle

%The ability to tune the electronic properties of a material using external stimuli is extremely useful in science and technology. Field effect transistors are an excellent illustration of this, where functionality is achieved by applying an electric field that controls the charge density of the device. The recent interest in manipulating spin for possible applications in quantum computing further demonstrate the advantages of wielding an electronic degree of freedom. 

In materials such as Si \cite{Ando.RevModPhys,Lai.PRL.2004,Goswami.NatPhys.2007,Lu.PRB.2012,Salfi.NatMater.2014}, AlAs \cite{Lay.APL,Etienne.Science,Ettiene.APL,Gunawan.Ballistic,ValleySkyrmions,ValleySusceptibility,Gunawan.PRB2006,Shayegan.Review,Bishop,Grayson.APL,Gokmen.NatPhys,Padmanabhan.PRL,Wegs1.PRB,Chung.PRM}, graphene \cite{Zhang.Nature.2005,Novoselov.Science.2007,Dean.NatPhys.2011,Feldman.PRL.2013,Gorbachev.Science.2014,Zibrov}, and MoS$_2$ \cite{Zeng.NatNano.2012,Xao.PRL.2012,Mak.Science.2014}, in addition to charge and spin there is yet another tunable electronic degree of freedom, namely their multiple conduction-band minima located away from the Brillouin zone center reciprocal space. Commonly denoted as `valleys', these conduction-band minima have an inherent degeneracy determined by the symmetry of the crystal structure, and perturbations that break this symmetry can be used to control the valley occupancy of electrons. Two-dimensional electron systems (2DESs) hosted in these multivalley materials are therefore remarkably versatile platforms that can be used to study new concepts for information processing as well as complex many-body interactions that involve multiple components \cite{Valleytronics.Review}. 

\begin{figure*} [t]
\centering
    \includegraphics[width=.95\textwidth]{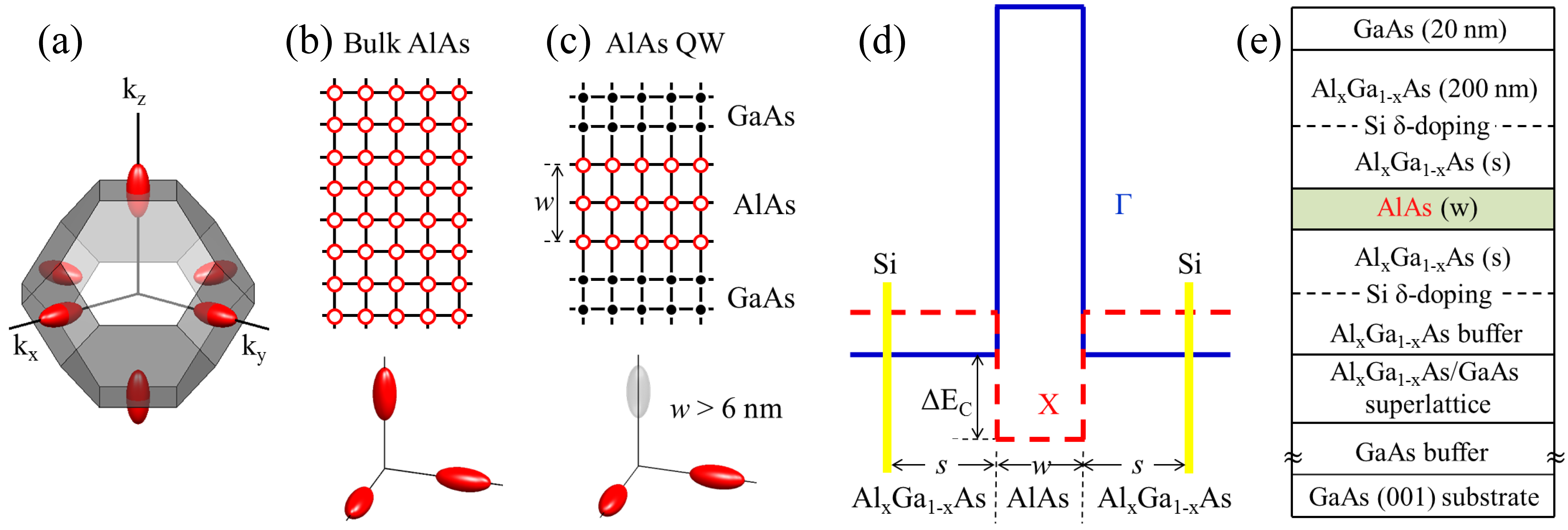}

%\begin{figure*} [t]
%  \begin{center}
%    \psfig{file=Fig1_r10.png, width=0.95\textwidth }
% \end{center}
 \caption{\label{fig1} (a) Schematic diagram of the first Brillouin zone in bulk AlAs. Three full ellipsoids (or equivalently, six half-filled ellipsoids) are occupied at the X points in reciprocal space. Electron occupation for (b) bulk AlAs and (c) AlAs QWs grown on a GaAs substrate. The biaxial compression from the GaAs substrate causes the two in-plane valleys to be occupied for AlAs QWs with well width $w>6$ nm. (d) A schematic conduction-band diagram of the samples used in our study. The Al$_{0.33}$Ga$_{0.67}$As barriers flanking the AlAs QW are $\delta$-doped with Si. The well width $w$ and spacer thickness $s$ are variables. (e) The sample structure for our AlAs quantum wells.}
\end{figure*}

A good example is the AlAs 2DES, where the prospect of valleytronics was first discussed in experiments demonstrating the possibility to tune valley occupancy via uniaxial strain \cite{ValleySusceptibility} or lateral confinement \cite{Gunawan.PRB2006}. It is also a system in which it is possible to observe quantum Hall ferromagnets \cite{Etienne.Science,ValleySkyrmions,Padmanabhan.PRL} and valley skyrmions \cite{ValleySkyrmions}, as well as composite fermions with a valley degree of freedom \cite{Bishop}. Moreover, the electrons in AlAs 2DESs have a large, anisotropic mass, making the system an intriguing candidate to investigate the physical properties of delicate many-body ground states \cite{Shayegan.Review,Gokmen.NatPhys}. 

Because of these exciting prospects, there has always been a strong motivation to improve the AlAs 2DES quality but it has been in a stalemate for over a decade now \cite{Ettiene.APL,Grayson.APL,Wegs1.PRB,Chung.PRM}. Here we report more than an order of magnitude improvement in the mobility of AlAs 2DESs, with a peak value of $\mu=2.4\times10^6$ cm$^2$V$^{-1}$s$^{-1}$ at a 2D electron density of $n=2.2\times10^{11}$ cm$^{-2}$. The high quality of our samples is further demonstrated by the observation of high-order fractional quantum Hall states up to the filling $\nu=8/17$ in a sample with $n=1.0\times10^{11}$ cm$^{-2}$. In a sample with $n=0.60\times10^{11}$ cm$^{-2}$, we can even observe a $\nu=1/5$ fractional quantum Hall state, flanked by reentrant insulating phases which are generally interpreted as pinned Wigner crystal states, a feat only achieved in the cleanest of 2DESs. These results represent tremendous advancements in AlAs 2DES quality, putting it in a regime comparable to that of state-of-the-art 2DESs.    

%The AlAs 2DESs used in our study are designed to have electrons occupy two in-plane valleys. 
Figure 1(a) shows a schematic diagram of the first Brillouin zone for bulk AlAs. Electrons in bulk AlAs occupy three full ellipsoids of revolution centered at the X points in reciprocal space, with each ellipsoid having a longitudinal effective mass $m_l=1.1$ and transverse effective mass $m_t=0.2$, both in units of free electron mass (see Fig. 1(b)). A 2DES can be hosted in an AlAs quantum well (QW) in a modulation-doped structure similar to that of a GaAs 2DES, with the primary difference being that the lowest energy band of the QW is the X band for AlAs while it is the $\Gamma$ band for GaAs \cite{Ettiene.APL,Shayegan.Review,Chung.PRM}. A representative conduction-band diagram for our structures in the vicinity of the AlAs QW is given in Fig. 1(d), while a detailed description of our sample structure is shown in Fig. 1(e). Because of the large mass anisotropy, when an AlAs QW is defined along a $<$100$>$ direction, one would expect that electrons occupy the out-of-plane valley since the confinement mass for this valley ($m_l$) is larger compared to the confinement mass for the two in-plane valleys ($m_t$). However, when an AlAs QW is grown pseudomorphically on a GaAs substrate by molecular beam epitaxy (MBE), the slightly larger lattice constant of AlAs puts the AlAs layer under biaxial compression. This lowers the energies of the two in-plane valleys so that these valleys are occupied when the AlAs QW width ($w$) is larger than $\sim6$ nm as shown in Fig. 1 (c) \cite{Shayegan.Review,AlAsNarrow1,AlAsNarrow2}. All of our AlAs QW samples presented here have $w>10$ nm, meaning we are studying AlAs 2DESs with two in-plane valleys occupied. 

%These AlAs 2DESs are hosted in a modulation-doped structure similar to that of a GaAs 2DES, with the primary difference being that the lowest energy band of the QW is the X band for AlAs while it is the $\Gamma$ band for GaAs \cite{Shayegan.Review,Chung.PRM}. A representative conduction-band diagram for our structures in the vicinity of the AlAs QW is given in Fig. 1(d).

%Although there are some reports on AlAs quantum wells with $w>10$ nm that study the 2DES with two in-plane valleys occupied \cite{Ettiene.APL,Grayson.APL,Wegs1.PRB}, none of them show AlAs quantum wells with $w\geq20$ nm. This implies that the quality of the 2DESs was suboptimal since the quantum wells are relatively narrow and hence more susceptible to the influence of layer fluctuations during MBE growth.

It is remarkable that the vast majority of AlAs QWs used to study 2DESs with two in-plane valleys occupied have $w\simeq10$ nm. This is at first sight surprising, as 2DESs in such narrow QWs are bound to suffer from interface roughness scattering induced by the layer-thickness fluctuations during the MBE growth. For a 10-nm-wide QW, even a monolayer fluctuation ($\simeq0.28$ nm for GaAs or AlAs) would cause the ground-state energy $E_0=\frac{\pi^{2}\hbar^2}{2m^{*}w^2}$ to fluctuate by $\simeq6\%$ (assuming an infinite potential well). Such fluctuations would lead to local density variations in the plane of the 2DES. Accordingly, an electron traversing this plane would experience scattering due to sporadic changes in its Fermi wavevector. When $w$ is increased, $E_0$ becomes less sensitive to layer fluctuations, and therefore we can expect less scattering and better quality. The fact that narrow QWs are detrimental to 2DES quality is well understood and has been studied in detail both experimentally and numerically for GaAs \cite{Sakaki.APL,Li.SST.2005,Kamburov.APL} and AlAs QWs \cite{Shayegan.Review,JAP.AlAsBulkImpurities}. 

Based on the above arguments, we can speculate that increasing the QW width to $w\geq20$ nm may yield higher-quality AlAs 2DESs. However, the MBE growth of AlAs QWs that fit this criterion is not so trivial because Al is an extremely reactive element. Increasing $w$ in an AlAs QW also increases the amount of background impurities near the AlAs 2DES, resulting in the possibility that the 2DES quality would actually worsen despite the discussion in the previous paragraph. The absence of any reports on AlAs QWs with $w\geq20$ nm strongly supports this hypothesis.

%Based on the above arguments, we can speculate that increasing the QW width to $w\geq20$ nm may yield higher-quality AlAs 2DESs. However the MBE growth of AlAs QWs that fit this criterion is not so trivial because Al is an extremely reactive element. This is well demonstrated by the fact that previously reported high quality AlAs quantum wells already have a background impurity level that is more than an order of magnitude higher than that of the Al$_x$Ga$_{1-x}$As barrier in typical  Al$_x$Ga$_{1-x}$As/GaAs heterostructures even when the AlAs quantum well width is only $\simeq 15$ nm \cite{JAP.AlAsBulkImpurities}. Under similar growth conditions, increasing $w$ would further increase the total amount of background impurities in the quantum well and there is a possibility that the 2DES quality actually worsens despite the discussion in the previous paragraph. The lack of any reports on AlAs quantum wells with $w\geq20$ nm in literature strongly supports this hypothesis.

Therefore, to properly verify whether the AlAs 2DES quality can be improved by increasing the QW width, it is crucial to first significantly decrease the amount of background impurities present in the AlAs itself. Our approach to this problem was to systematically purify our Al source in the ultra-high-vacuum environment of our MBE chamber. As described in detail elsewhere \cite{Chung.PRM2}, we measured the mobility of specially designed GaAs 2DESs as we outgassed our Al source and used the mobility to evaluate the purity of the Al cell. The main idea was to exploit the fact that impurities surface segregate in the MBE growth direction for Al$_x$Ga$_{1-x}$As. If an Al$_x$Ga$_{1-x}$As barrier lies directly below a GaAs 2DES, any impurities segregating at the MBE surface of the underlying Al$_x$Ga$_{1-x}$As barrier will cause the mobility of the GaAs 2DES to decrease. Assuming that the vacuum conditions are unchanged, the total amount of surface segregated impurities should depend on the purity of the Al source and the thickness of the underlying Al$_x$Ga$_{1-x}$As barrier. Indeed, as we outgassed our Al source we observed the GaAs 2DES mobility increase and approach values in hetero-structure samples which have no underlying Al$_x$Ga$_{1-x}$As barrier. After sufficiently purifying our Al source, we were able to grow a GaAs QW that had no degradation in mobility from the segregating impurities even with a 350-nm-thick underlying Al$_{0.33}$Ga$_{0.67}$As barrier. This result roughly implies that there would not be a significant amount of impurities even in a $350\times0.33\simeq116$ nm thick AlAs QW, which leaves plenty of room to test out AlAs 2DESs with $w\geq20$ nm.

\begin{figure}[t]
 
 \centering
    \includegraphics[width=.45\textwidth]{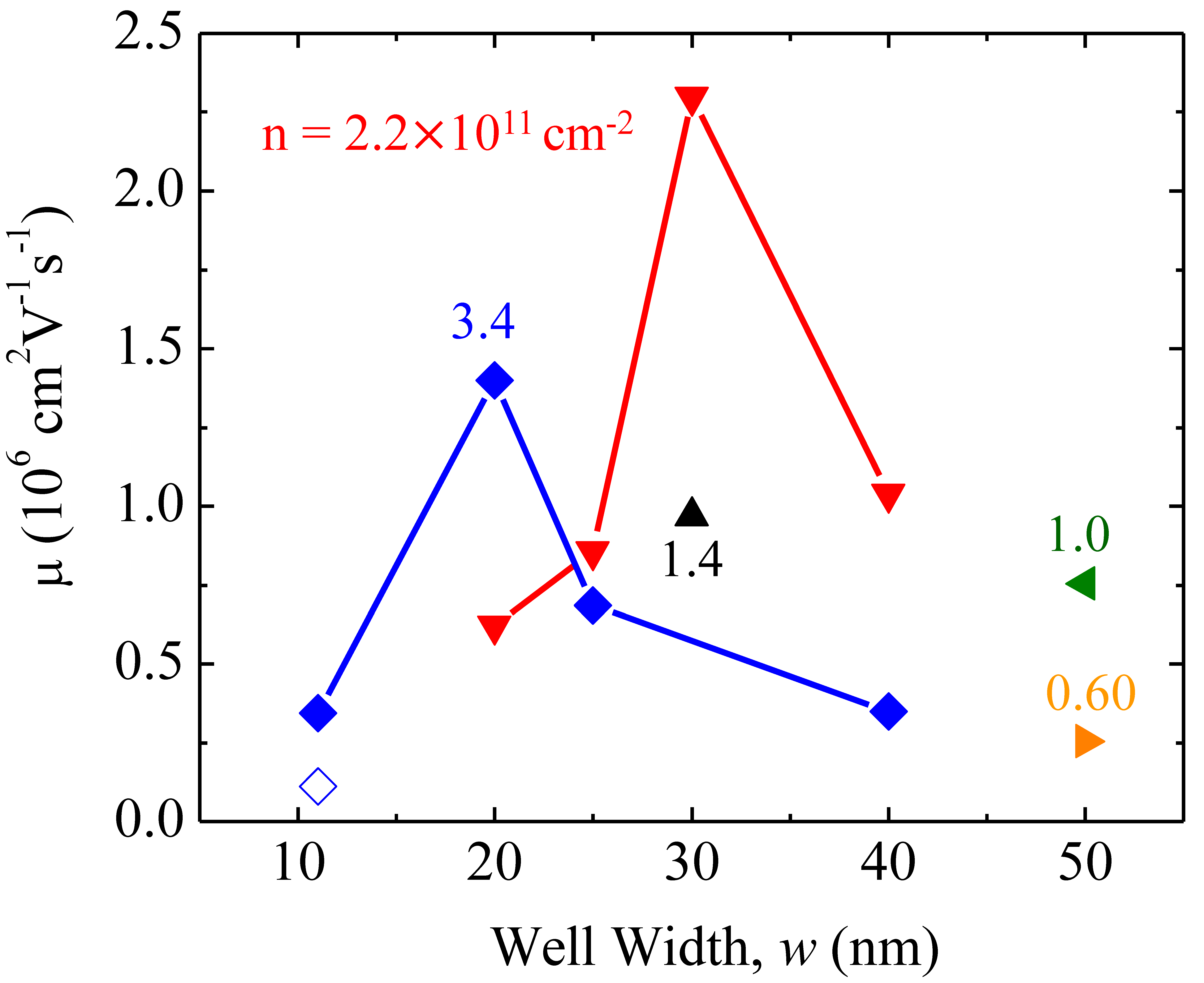} 
% \begin{center}
%    \psfig{file=Fig2_r9.png, width=0.45\textwidth }
%  \end{center}
  \caption{\label{fig2} Measured mobilities of our AlAs QWs (closed symbols) at $T\simeq0.3$ K as a function of QW width. Each color represents a data set from a series of samples grown to have the same density where $w$ was the only variable. The density was varied by changing the spacer thickness $s$ in the Al$_{0.33}$Ga$_{0.67}$As barrier: $s=60$, 100, 150, 220, and 350 nm for the densities $n=3.4$, 2.2, 1.4, 1.0, and $0.60\times10^{11}$ cm$^{-2}$, respectively. The open symbol represents a data point from Ref. \cite{Ettiene.APL}.}
\end{figure} 
\begin{figure}[t]

\centering
    \includegraphics[width=.45\textwidth]{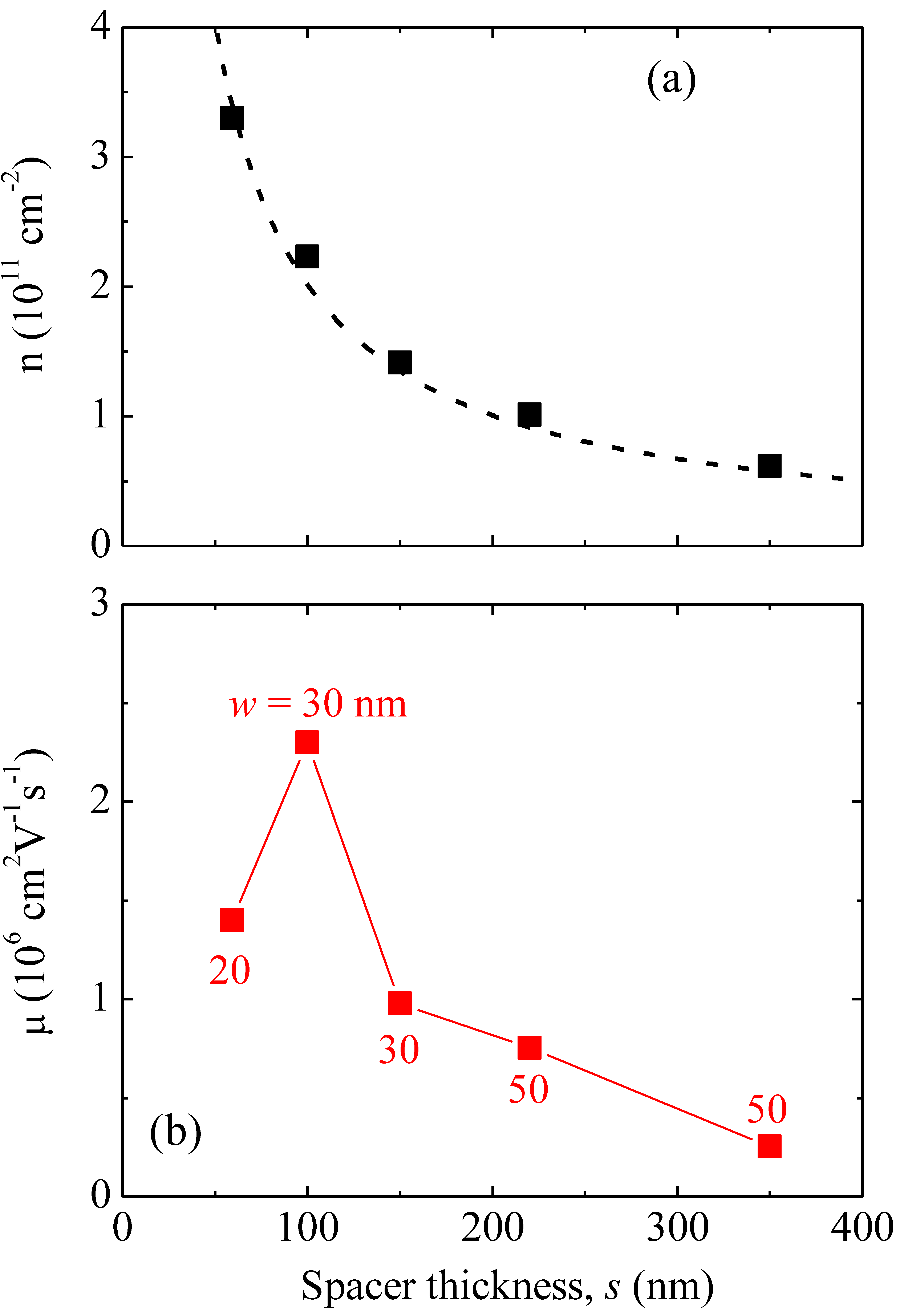} 
%  \begin{center}
 %   \psfig{file=Fig3_r9.png, width=0.45\textwidth }
 % \end{center}
  \caption{\label{fig3} (a) Density, and (b) mobilities, of our AlAs 2DESs as a function of spacer thickness. The dashed line in (a) has the form $n\propto s^{-1}$ and is a fit to the data points. In (b), the QW width $w$ is given for each data point.}
\end{figure} 

For our study, modulation-doped AlAs QWs were grown on $<$001$>$ GaAs substrates by molecular beam epitaxy. The main structure is comprised of an AlAs QW flanked on both sides by Al$_{0.33}$Ga$_{0.67}$As barriers that were $\delta$-doped with Si at a concentration of $\sim6\times10^{11}$ cm$^{-2}$ (see Fig. 1(e)). The growth temperature was 645\textdegree C except during Si $\delta$-doping, where the temperature was decreased to 500\textdegree C. Growth rates for the Ga and Al cells were calibrated by reflection high-energy electron diffraction oscillations and set to be 0.28 nm/s and 0.14 nm/s, respectively. The As beam flux was fixed to $5.8\times10^{-6}$ Torr, and the Ga:As flux ratio was 1:13 while the Al:As ratio was 1:20. The QW width was varied to optimize the 2DES quality and the spacer was varied to control the 2DES density. Electrical measurements were performed on $\simeq$ 4 mm $\times$ 4 mm van der Pauw samples using the low-frequency lock-in technique at two temperatures, 300 mK to evaluate the mobility of the samples and 100 mK to measure magneto-transport traces. A red light-emitting diode (LED) was illuminated for $\sim5$ minutes at 10 K during the cooldown procedure to achieve maximum carrier density through persistent photo-conductivity \cite{Chung.PRM}, and after the red LED was turned off the samples were then cooled to the base temperature. The cooldown procedure was fixed for all measurements.

Figure 2 shows the measured 2DES mobility of our AlAs QWs as a function of $w$ for various 2DES densities ($n$). We varied $n$ by changing the spacer thickness $s$ as described in the caption of Fig. 2. Comparing the mobility of the sample with $n=3.4\times10^{11}$ cm$^{-2}$ and $w=11$ nm with previous results \cite{Ettiene.APL}, the effect of our Al source purification is already evident. It is also clear from Fig. 2 that having a QW with $w\geq20$ nm is instrumental in achieving optimal quality in AlAs 2DESs, where mobilities exceeding $10^6$ cm$^2$V$^{-1}$s$^{-1}$ are readily observed at different densities. At $n=2.2\times10^{11}$ cm$^{-2}$, the mobility peaks at $2.4\times10^6$ cm$^2$V$^{-1}$s$^{-1}$ for $w=30$ nm, more than ten times higher than previous results at similar densities \cite{Lay.APL,Ettiene.APL,Shayegan.Review,Grayson.APL,Chung.PRM}. Moreover, we were able to grow high-quality AlAs QWs as wide as 50 nm, in which we measure a mobility of $2.5\times10^5$ cm$^2$V$^{-1}$s$^{-1}$  even at the extremely low density of $0.60\times10^{11}$ cm$^{-2}$.  

The decrease in mobility observed in Fig. 2 at densities 3.4 and $2.2\times10^{11}$ cm$^{-2}$ as $w$ increases is attributed to scattering contributions from the second subband. This is consistent with the fact that the mobility peaks at a smaller $w$ for the higher density samples. Preliminary results deduced from our Schr\"{o}dinger-Poisson solver also support this view. To ensure that no additional scattering from the second subband ensues in our lower density samples, $w$ was fixed to 30 nm for the sample with $n=1.4\times10^{11}$ cm$^{-2}$. The widest well width of $w=50$ nm was only used for the two substantially more dilute cases with $n=1.0\times10^{11}$ cm$^{-2}$ and $0.60\times10^{11}$ cm$^{-2}$. 

%The decrease in mobility observed in Fig.2 at the densities 3.4 and $2.2\times10^{11}$ cm$^{-2}$ as $w$ increases is attributed to scattering contributions from the second subband. This is consistent with the fact that the mobility peaks at different well widths for the two densities. Assuming that the two in-plane valleys are in exact balance, using a self-consistent Schr\"{o}dinger-Poisson solver we deduce that the second sub-band should be occupied at $w\simeq32$ nm and $\simeq39$ nm for the densities of 3.4 and $2.2\times10^{11}$ cm$^{-2}$ respectively. Considering that some residual symmetry breaking strain from growth, processing, or the cooldown procedure could be present in the sample and cause the AlAs 2DES to be off balance, these numbers are reasonably in agreement with the data. From our Schr\"{o}dinger-Poisson solver we expect that at a well width of $w=50$ nm the second sub-band starts to fill up at a density of $\sim1.5\times10^{11}$ cm$^{-2}$ given the two valleys are degenerate. To ensure that no additional scattering derives from the second sub-band in our lower density samples, $w$ was fixed to 35 nm for the sample with $n=1.4\times10^{11}$ cm$^{-2}$ and the widest well width of $w=50$ nm was only used for the two substantially more dilute cases with $n=1.0\times10^{11}$ cm$^{-2}$ and $0.6\times10^{11}$ cm$^{-2}$. 

\begin{figure*}[t]
\centering
    \includegraphics[width=.95\textwidth]{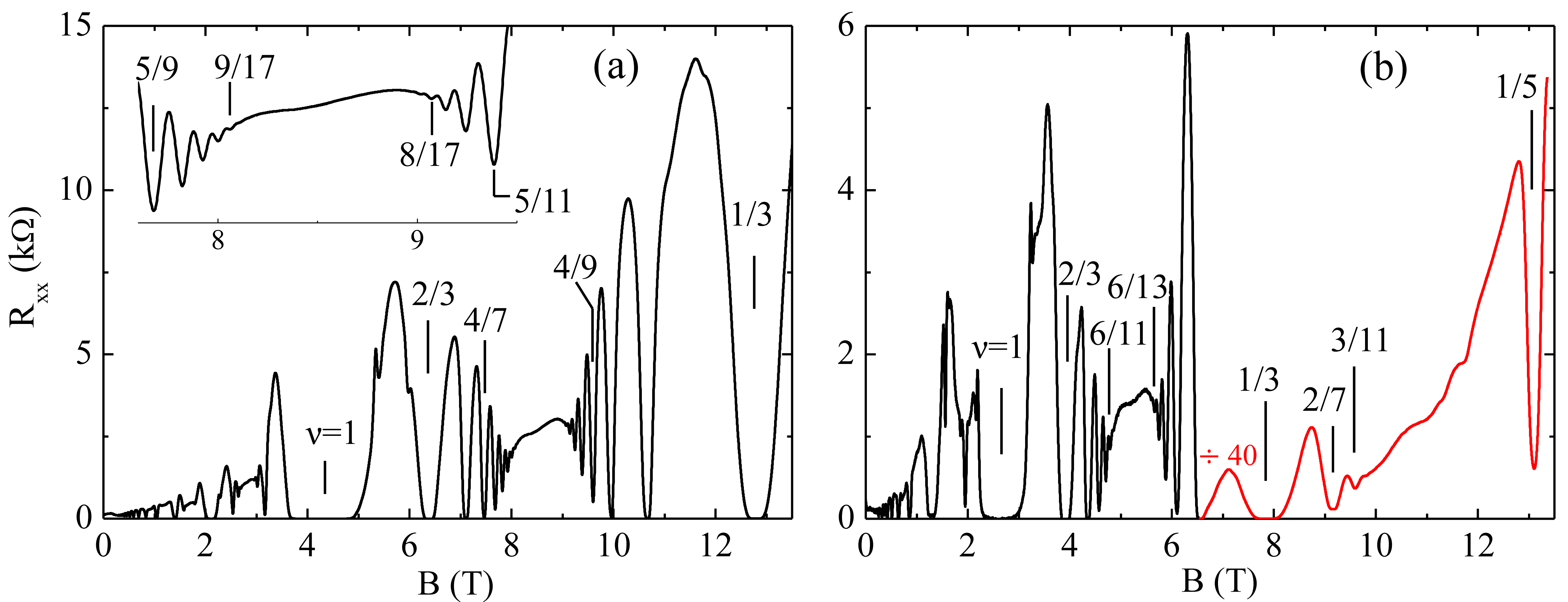} 

 % \begin{center}
  %  \psfig{ file=Fig4_r8.png, width=0.95\textwidth }
  %\end{center}
  \caption{\label{fig4} Magneto-transport data of our AlAs samples with 2DES densities of (a) $n=1.0\times10^{11}$ cm$^{-2}$ ($w=50$ nm, $\mu=7.5\times10^5$ cm$^{2}$V$^{-1}$s$^{-1}$) and (b) $n=0.6\times10^{11}$ cm$^{-2}$ ($w=50$ nm, $\mu=2.5\times10^5$ cm$^{2}$V$^{-1}$s$^{-1}$), measured at $T=100$ mK. The data shown in red in (b) was divided by 40 to match the scale shown in the y-axis. The inset in (a) enlarges the region near $\nu=1/2$ to show the higher-order FQHE states more clearly.}
\end{figure*}

Figure 3 shows the density and mobilities of our samples as a function of spacer thickness. As expected for modulation-doped structures, the density of our 2DESs follows the $n\propto\,s^{-1}$ rule, which is depicted by the dashed line in Fig. 3(a). The variation of the mobility in Fig. 3(b) is not monotonous, peaking at $s=100$ nm and then decreasing as the spacer thickness increases and the density decreases. Considering that the enhanced screening with larger $n$ usually results in the mobility to also increase in a 2DES, the behavior of mobility in Fig. 3(b) when $s\leq100$ nm seems puzzling. There are two possible causes for the drop in mobility as $s$ decreases well below 100 nm. One is the increased scattering from the intentional remote dopants and the other is the inevitably narrow QW width required at the higher 2DES density to ensure that the upper electric subband is not occupied. Comparing the mobilities of samples with $w=20$ nm for $n=3.4$ and $2\times10^{11}$ cm$^{-2}$ in Fig. 2, it is clear that the former is not the main cause. The mobilities in Fig. 3(b) therefore include not only the effect of change in 2DES density from varying $s$, but also the associated change in the representative QW width. The data imply that at small $s$ the increased scattering from the narrower QW width overcomes the effect of higher 2DES density while at larger $s$ the QW width can be wide enough that only the effect of smaller 2DES density is evident. Thus the results shown here provide a general guideline for AlAs 2DES quality optimization when both QW width $w$ and 2DES density $n$ are variables. 

%the interplay the between 2DES density and the QW width. For example, according to the data in Fig. 2, the samples with the smallest spacer thickness of $s=60$ nm have a 2DES density where the condition $w\leq20$ nm must be met to prevent any scattering from the second subband. However, considering the data for $n=2.2\times10^{11}$ cm$^{-2}$ in Fig. 2, it is clear that the relatively narrow QWs ($w\leq25$ nm) significantly limit the mobility.

It is interesting to compare the mobilities of our AlAs 2DESs with those of high-quality GaAs 2DESs. Because the two systems have distinct electrical parameters (effective mass, Fermi sea anisotropy, etc.), it is impossible to compare their mobilities directly. However, given a situation where both 2DESs have the same density and similar structural conditions such as spacer thickness and QW well width, we can make a reasonable comparison. An exceptionally high-quality GaAs 2DES that exhibits the $\nu=1/7$ fractional quantum Hall effect (FQHE) has been reported to have a mobility of $10\times10^6$ cm$^2$V$^{-1}$s$^{-1}$ at $n=1.0\times10^{11}$ cm$^{-2}$ \cite{Pan.PRL}. The sample in Ref. \cite{Pan.PRL} has a QW with $w=50$ nm and a spacer thickness $s=220$ nm, which are the exact structural paramaters of our AlAs 2DES with the same density. Assuming that the mobility scales inversely with the effective mass ($m^*$) via $\mu=\frac{e\tau}{m^*}$ where $e$ is the electron charge and $\tau$ is the scattering time, we can compare our results with Ref. \cite{Pan.PRL}. For transport along [100], and assuming that the two valleys in AlAs are equally occupied, when we scale our AlAs 2DES (effective transport mass $m^*=\frac{2 m_l m_t}{(m_l+m_t)}=0.34$) to a GaAs 2DES ($m^*=0.067$), we obtain a mobility of $3.8\times10^6$ cm$^2$V$^{-1}$s$^{-1}$. Although there are other subtle differences between the two systems such as barrier composition and valley occupancy, this rescaled mobility suggests that our AlAs 2DESs are approaching the quality of ultra-clean GaAs 2DESs.

%\red{At first sight the fact that we obtain a $\mu\propto\,n^{1.65}$ relation by fitting the data points with $s\geq100$ nm as shown in the inset of Fig. 3(b) seems to support the view that the effect of decreasing 2DES density governs the mobility in samples with $w\geqq30$ nm. However, considering that in some of these samples $s$ is quite large (up to 350 nm), it is unreasonable to assume that remote ionized impurities dominate scattering in all cases \cite{Jiang.APL.1988}. We are unsure about the exact physics that can explain this power law relation at the moment, although it could be an artifact of the decrease in 2DES density in combination with effects such as interface scattering, valley occupancy, etc.}

%The latter argument is further corroborated by the fact that we obtain a $\mu\propto\,n^{1.65}$ relation by fitting the data points with $s\geqq100$ nm as shown in the inset of Fig. 3(b). 

Given the high mobility of our AlAs 2DESs, we also performed magneto-transport measurements at low temperatures to further attest their unprecedented quality. Figures 4(a) and (b) show the longitudinal resistance ($R_{xx}$) traces taken at $T=100$ mK for our samples with 2DES densities of $n=1.0$ and $0.60\times10^{11}$ cm$^{-2}$, respectively. In comparison with previous results \cite{Lay.APL,Ettiene.APL,Shayegan.Review,Grayson.APL,Chung.PRM}, it is clear that our samples have significantly better quality as demonstrated from the observation of high-order FQHE features up to the Landau level filling of  $\nu=8/17$ at the density of $n=1.0\times10^{11}$ cm$^{-2}$. Moreover, as shown in Fig. 4(b), in our lowest-density sample with $n=0.60\times10^{11}$ cm$^{-2}$ we observe the delicate $\nu=1/5$ FQHE, flanked by insulating phases which likely signal pinned Wigner crystal states \cite{Wigner1,Wigner2}. Considering that the $\nu=1/5$ state has only ever been reported in the cleanest of 2D carrier systems \cite{Feldman.PRL.2013,Wigner1,Wigner2,Santos.PRB.1992}, this spectacular result highlights the extraordinary quality of our AlAs 2DESs. 

It is worthwhile to compare our results to those for other multivalley 2DESs such as Si or monolayer graphene. In a Si/SiGe field-effect transistor structure, it has been reported that Si 2DESs which occupy two out-of-plane valleys can achieve a mobility of $1.6\times10^6$ cm$^{2}$V$^{-1}$s$^{-1}$ at $n\simeq2.6\times10^{11}$ cm$^{-2}$ and $T=30$ mK \cite{Lu.PRB.2012}. Since the transport effective mass for this case ($m^*=m_t=0.19$) is smaller than our AlAs 2DESs, we expect that our samples have higher quality. This conjecture is consistent with the fact that while the Si 2DES data in Ref. \cite{Lu.PRB.2012} show high-order FQHE minima up to $\nu=4/9$ in the extreme quantum limit, our samples exhibit FQHE features up to $\nu=8/17$ even at lower densities (Fig. 4(a)). As for monolayer graphene, a direct comparison to our AlAs data is difficult because $m^*$ in graphene is ill defined and mobility values are seldom reported \cite{Dean.NatPhys.2011}. Moreover, rather than magneto-transport traces at a fixed density, often compressiblity data, measured at a fixed magnetic field and as a function of varying the density, are presented \cite{Feldman.PRL.2013,Zibrov}. Nevertheless, a comparison of the high-order FQHE features observed in the two systems is illuminating. The highest-order FQHE feature reported for graphene is at $\nu=6/13$ at a magnetic field of $\simeq28$ T \cite{Zibrov} while we observe FQHE features at even larger denominator fillings at magnetic fields which are a factor of three smaller (Fig. 4(a)). 

%As for monolayer graphene, the low temperature mobilities have been reported to be $\sim10^5$ cm$^{2}$V$^{-1}$s$^{-1}$ when $n\sim10^{11}$ cm$^{-2}$ \cite{Dean.NatPhys.2011}. Although it is difficult to compare the magneto-transport data of this case to our own because $n$ is a variable in the longitudinal resistance traces of Ref. \cite{Dean.NatPhys.2011}, looking at the extreme quantum limit where $n$ is small enough to compare to the range of ours, it is clear that many more fractional quantum Hall minima are observed in our data. The $R_{xx}$ traces shown in Fig. 4 are remarkable considering they show even more fractional quantum Hall features than when local compressibility measurements are performed on suspended monolayer graphene \cite{Feldman.PRL.2013}. In addition to the high quality of our AlAs 2DESs in comparison with other multivalley systems, we reiterate that AlAs 2DESs differentiate from Si or monolayer graphene in the aspect that the valley occupancy can be externally tuned using strain, which could be extremely useful in the study of many-body interactions or other practical applications. 

We emphasize that the high-quality samples presented in our report were achieved by purifying our Al source and subsequently increasing and optimizing the AlAs QW. Our results put AlAs 2DESs in a new league with quality comparable to the finest 2D carrier systems currently available. We add that, besides their extremely high quality, our AlAs 2DESs differentiate from other multivalley materials such as monolayer graphene in that the valley occupancy can be externally tuned using strain \cite{Shayegan.Review}. This tunability is extremely useful in studies of many-body interactions or other practical applications

\begin{acknowledgments}
We acknowledge support through the NSF (Grants DMR 1709076 and ECCS 1508925) for measurements, and the NSF (Grant MRSEC DMR 1420541), the Gordon and Betty Moore Foundation (Grant GBMF4420), and the Department of Energy Basic Energy Sciences (DEFG02-00-ER45841) for sample fabrication and characterization.
 \end{acknowledgments}

\end{document}